\documentclass[aps,prb,twocolumn,showpacs,a4paper,superscriptaddress]{revtex4-1}
\usepackage{graphicx,bm,amsmath,dcolumn,amssymb}
\usepackage{longtable}
\graphicspath{{figures}}
\usepackage{geometry,hyperref,color}
\begin{document}
\title{Dynamical relaxation behaviors of a critical quench}
  \author{Yin-Tao Zou}
  \affiliation{School of Microelectronics $\&$ Data Science, Anhui University of Technology, Maanshan 243002, China }
   \author{Chengxiang Ding}
  \email{dingcx@ahut.edu.cn}
   \affiliation{School of Microelectronics $\&$ Data Science, Anhui University of Technology, Maanshan 243002, China }
\date{\today}
\begin{abstract} 
  We study the universal dynamical relaxation behaviors  of a quantum XY chain following a  quench, paying special attention to the case that the  prequenched Hamiltonian, or the postquenched Hamiltonian, or both of them are at critical points of  equilibrium quantum phase transitions. In such ``critical quench", we find very interesting real-time dynamical scaling behaviors and the crossover phenomena between them. 
  For a quench from a noncritical point to  a critical point, we find that, compared to the noncritical quench, the universal power-law scaling behavior  does not change;  
  however, there may be a crossover between the exponential decaying behavior and the power-law scaling.
  For a quench from a critical point to a noncritical point, the power-law scaling behaviors $t^{-3/2}$ and $t^{-3/4}$ in the noncritical quenches may be changed to $t^{-1}$ and $t^{-1/2}$, respectively.  
 If the prequenched Hamiltonian is set to be a point that is close to but not exactly at a critical point, we find interesting crossover phenomena between different power-law scaling behaviors.  
  We also study the quench from the vicinity of a multicritical point,  we find crossover behaviors that are induced by a different mechanism, and  new crossover exponent is found. 
  All the results are related to the gap-closing properties of the energy spectrum of the critical points.
  
 
\end{abstract}
\maketitle 

\section{Introduction}


The study of the nonequilibrium properties of a many-body quantum system is a hot topic in the condensed matter physics.
The research interest include both the properties of the steady state in long-time limit  and also the properties of  dynamical  relaxation. 

A lot of important questions related to the properties of steady state have been extensively studied, 
such as the question of thermalization\cite{ETH}, the thermodynamic relations\cite{TDR}, the zero and second laws\cite{zero}, the heat conduction\cite{heatcod}, the thermodynamic uncertainty relation\cite{TDU}, 
the thermodynamic force\cite{TDF}, the first-type of dynamical quantum phase transition (DQPT-I)\cite{DQPT-I,DQPT-Ia}, and so forth\cite{QI,QI2,Porta2018}. 
The properties of the dynamical behaviors have also been widely studied in different aspects, such as the Kibble-Zurek mechanism\cite{Kibble1976,Kibble1980,Zurek1985,Zurek1996}, the entropy production\cite{ent}, the heating process\cite{heat}, the aging phenomena\cite{aging}, the dynamical topology\cite{Dtop1,Dtop2}, the second type of dynamical quantum phase transition (DQPT-II)\cite{Heyl2013,DQPT-IIx}, and so forth\cite{TranD,TanD2,You2022}.

In the study of the aforementioned questions,  an important concept, the universality, which originates from the study of the equilibrium phase transition,  has been applied in part of them, 
such as the Kibble-Zurek mechanism and the DQPTs.  In DQPT-I,   the critical behavior is revealed by the later-time average of the order parameter, it is the generalization of the equilibrium state phase transition to the nonequilibrim system. The concept of universality, i.e., the definition of critical exponents of such nonequilibrium phase transition is very similar to the equilibrium  phase transition.
The DQPT-II, which is closely related to the DQPT-I,  is defined on the  phase-transition-like nonanalytic behaviors of the so-called dynamical free energy at a series of critical times when the overlap between the initial state and the evolving state is zero\cite{Heyl2013}.  This type of DQPT has also been extensively studied in different quantum systems with different generalizations,  including the mixed state\cite{mix1,mix2},  the open system\cite{open}, the Floquet system\cite{floquet}, and so forth\cite{ding2020,exited}.  The concept of universality has also been applied to such type of phase transition\cite{univ1}, although the theory is still in developing\cite{univ2,univ3,Khan2023}.  

Interestingly, besides the aforementioned DQPTs, there is a third type of DQPT (DQPT-III)\cite{DQPT-3,DQPT-3a,pbc}, which concerns the asymptotic behavior of a physical variable approaching the steady value in the long-time limit. It is shown that  the difference between the value of the correlation function  at time $t$ from the value at steady state may scales as $t^{-\mu}$, and the DQPT-III  is defined by the change of the scaling exponent $\mu$, which has been found to happen at a critical frequency of a periodically driven integrable system\cite{DQPT-3,DQPT-3a,pbc}.   
Here it should be emphasized that, no matter whether there is a DQPT, the power-law decaying behavior  of the correlator is also  a universal property of the dynamics, 
i.e., the decaying exponent $\mu$  is determined by what phase the system quench to, not depending on the details of system.
 Such type of dynamical universality has also been studied in the  aperiodically driven systems\cite{apbc},  the stochastic driven system\cite{stoch}, and the noise driven system\cite{noise}. 
 
 Recently, such type of universal behaviors have  also been found in the quench dynamics of both the noninteracting and interacting integrable systems\cite{quench1,quench2}, with different scaling exponents. 
 In a quantum XY chain\cite{quench1}, Makki and coauthers find the $t^{-1/2}$, the $t^{-3/2}$, and the $t^{-3/4}$ power laws of decaying, depending on the postquench Hamiltonian is in the commensurate phase, the incommensurate phase, or the boundary between the two phases, respectively.  
 More importantly, it is theoretically proved  by the stationary phase approximation (SPA) that 
 the scaling law depends on the structure of the energy spectrum of the postqenched Hamiltonian, which deeply reveals the nature of such type of dynamical universality. 
  
 In the current paper, we study the  universal relaxation behaviors of a quantum XY chain following a quench, paying special attention to the case that the  prequenched Hamiltonian, or the postquenched Hamiltonian, or both of them are at critical points of  equilibrium quantum phase transitions, which is dubbed ``critical quench".  We find very interesting dynamical relaxation behaviors in such type of quench, the gap-closing of the energy spectrum may change the power-law scaling behaviors and lead to very interesting crossover phenomena. 
 Our study also includes the multicritical points, where the crossover phenomena are induced by a different mechanism, and new crossover exponent is found.

The paper is arranged as follows: In Sec. \ref{models} and Sec. \ref{method}, we introduce the model and  method, respectively; 
in Sec. \ref{sec_short}, we give the results of the critical quench, including the quench to a critical point,  the quench from a critical point, and the quench from the vicinity of a multicritical point; we conclude our paper in Sec. \ref{con}.  
\section{Models}
\label{models}
The model we studied is a quantum XY spin chain\cite{pfeuty1970,begin}, 
\begin{eqnarray}
	H&=&-\sum\limits_{j=1}^L\Big[\frac{1+\chi}{2}\sigma_j^x\sigma_{j+1}^x+\frac{1-\chi}{2}\sigma_j^y\sigma_{j+1}^y-h\sigma_j^z\Big],\label{shortXY}
\end{eqnarray}
where $\sigma^x,\sigma^y$, and $\sigma^z$ are the Pauli matrixes of spin 1/2, and $L$ the number of sites of the chain, in which the periodic boundary condition is applied.  
 
By the Jordan-Winger transformation, the model can be transformed to a free-fermion model
\begin{eqnarray}
	H_{f}&=&-\sum\limits_{j=1}^L(c_j^\dagger c_{j+1} +\chi c_j^\dagger c_{j+1}^\dagger+{\rm H.c.})\nonumber\\
	&&+h\sum\limits_{j=1}^L(2c_j^\dagger c_j-1), \label{shortfermion}
\end{eqnarray}
here, we have restricted our study in the even-fermionic-number-parity sector and adopted the antiperiodic boundary conditions.

By the Fourier transformation, the model can be transformed to the momentum space, 
\begin{eqnarray}
	H=\sum\limits_{k>0}H_k=\sum_{k>0}\mathbf{\Psi}_k^\dagger \mathbf{H}_k\mathbf{\Psi}_k
\end{eqnarray}
where  $\mathbf{\Psi}_k=(c_k,c^\dagger_{-k})^T$ and 
\begin{eqnarray}
	\mathbf{H}_k=
	\begin{pmatrix}
		z_k & -iy_k\\
		iy_k & -z_k
	\end{pmatrix}.
\end{eqnarray}
The wave vector $k$ belongs to $\{  \pm(2n-1)\pi/L, n=1, 2, \cdots, L/2 \}$ because we have applied the antiperiodic boundary conditions.
Here $z_k=2(h-\cos k)$ and $y_k=2\chi \sin k$.

\begin{figure}[htpb]
 	\centering
 	\includegraphics[width=0.9\columnwidth]{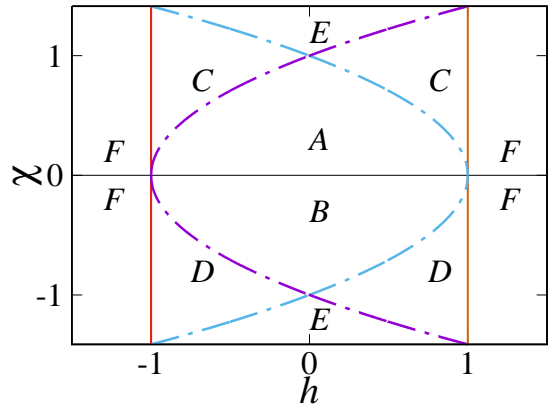}
 	\caption{Equilibrium phase diagram of the XY model (\ref{shortXY}): for $|h|<1$, the system is in the ferromagnetic or antiferromagnetic ordered phase; for $|h|>1$, the system is in the paramagnetic disordered phase;  A, B, and E are incommensurate phases; C, D, and F are commensurate phases; the dashed lines are the boundaries between the incommensurate and commensurate phases. }
 	\label{pd}
 \end{figure}

The Hamiltonian $H_k$ is already in a small Hilbert space of $2\times 2$, it can be easily diagonalized  by the Bogoliubov transformation 
\begin{eqnarray}
\gamma_k&=&u^*_kc_k+v^*_kc_{-k}^\dagger,\\
\gamma_{-k}^\dagger&=&-v_kc_k+u_kc_{-k}^\dagger,
\end{eqnarray}
where $u_k=\cos\theta_k, v_k=i\sin\theta_k$, with $\theta_k$ the Bogoliubov angle defined as 
\begin{eqnarray}
\tan\theta_k=y_k/z_k. \label{theta}
\end{eqnarray}
This gives the energy spectrum and the ground state of the model
\begin{eqnarray}
	\varepsilon_k&=&\sqrt{y_k^2+z_k^2},\\
	|\Phi\rangle&=&\prod\limits_{k>0}(u_k+v_kc_k^\dagger c_{-k}^\dagger|0\rangle),
	\end{eqnarray}
where $|0\rangle$ is the fermionic vacuum.  It is obvious that $k=0$ and $\pi$ are two gap-closing points of the model,  

The equilibrium phase diagram of the  model is shown in Fig.  \ref{pd}, where $h=1$ and $h=-1$  are the two critical lines determined  by the gap-closing momentum  $k=0$ and $\pi$, respectively;  the dashed lines are the boundaries between the commensurate and incommensurate phases,  which are  determined by $\cos k_0=h/(1-\chi^2)$, with $k_0=0$ and $\pi$.  The incommensurate phase is defined if there is an additional saddle point besides the saddle points $k=0$ and $\pi$; otherwise, it is a commensurate phase.

\section{Method}
\label{method}
The evolution of a quantum state $|\Phi(t)\rangle$ 
is determined by the  time-dependent Schr\"{o}dinger equation, for the quantum  XY model,  this is equivalent to solving the time-dependent  Bogoliubov–de Gennes (BdG) equation\cite{begin}
\begin{eqnarray}
	i\partial_t\phi_k(t)=\mathbf{H}_k\phi_k(t)
\end{eqnarray}
where $\phi_k(t)=(v_k(t), u_k(t)  )^T$. For the question of a quantum quench, 
the solution is a unitary evolution
\begin{eqnarray}
	\phi_k(t)=e^{-i\mathbf{H}_kt}\phi_k(0),\label{uniev}
\end{eqnarray}
where $\phi_k(0)=(v_k(0), u_k(0))^T$.   Using the Taylor expansion and the fact that $\mathbf{H}_k^2=\varepsilon_k^2\mathbf{I}$, where $\mathbf{I}$ is a $2\times2$ identity matrix, the calculation of Eq. (\ref{uniev}) can be easily performed, which gives the relation between $\phi_k(t)$ and $\phi_k(0)$, i.e., 
\begin{eqnarray}
	v_k(t)&=&v_k(0)[\cos(\varepsilon_kt)-i\sin(\varepsilon_kt)\cos\theta_k]\nonumber\\
	&&-u_k(0)\sin(\varepsilon_kt)\sin\theta_k\\
	u_k(t)&=&u_k(0)[\cos(\varepsilon_kt)+i\sin(\varepsilon_kt)\cos\theta_k]\nonumber\\
	&&+v_k(0)\sin(\varepsilon_kt)\sin\theta_k.
\end{eqnarray}
Using the expression of $v_k(t)$, the correlation  $C_{mn}(t)=\langle c_m^\dagger c_n\rangle$ can be calculated, which is 
\begin{eqnarray}
	C_{mn}(t)&=&\frac{1}{\pi}\int_{0}^{\pi}|v_k(t)|^2\cos[k(m-n)]\nonumber\\
	&=&C_{mn}(\infty)+\delta C_{mn}(t),
\end{eqnarray}
where 
\begin{eqnarray}
	C_{mn}(\infty)=&&\frac{1}{\pi}\int_{0}^{\pi}dk\Big[1-\cos\zeta_k\cos^2\theta_k\nonumber\\
	&&-\frac{1}{2}\sin\zeta_k\sin(2\theta_k)\Big]\cos[k(m-n)]
\end{eqnarray}
 is the value  of $C_{mn}(t)$ in the  steady state. Here $\zeta_k$ is the Bogoliubov angle of the prequenched Hamiltonian, $\theta_k$ is the Bogoliubov angle of the postquenced Hamiltonian.  $\delta C_{mn}(t)$ is the difference  between $C_{mn}(t)$ and $C_{mn}(\infty)$,  
 \begin{eqnarray}
 	\delta C_{mn}(t)=\delta C_{mn}^{(1)}(t)+\delta C_{mn}^{(2)}(t),
 \end{eqnarray}
where 
 	\begin{flalign}
&	\delta C_{mn}^{(1)}(t)\nonumber\\
&  =-\frac{1}{\pi}\int_{0}^{\pi}dk\cos\zeta_k\sin^2\theta_k\cos(2\varepsilon_kt)\cos[k(m-n)]\nonumber\\
& ={\rm Re}\Big\{-\frac{1}{\pi}\int_{0}^{\pi}dk\cos\zeta_k\sin^2\theta_ke^{2i\varepsilon_kt}\cos[k(m-n)]\Big\}, \label{int1}\\
&	\delta C_{mn}^{(2)}(t)\nonumber\\
& =\frac{1}{2\pi}\int_{0}^{\pi}dk\sin\zeta_k\sin(2\theta_k)\cos(2\varepsilon_kt)\cos[k(m-n)]\nonumber\\
& ={\rm Re}\Big\{\frac{1}{2\pi}\int_{0}^{\pi}dk\sin\zeta_k\sin(2\theta_k)e^{2i\varepsilon_kt}\cos[k(m-n)]\Big\}. \label{int2}
\end{flalign}

The asymptotic behavior of $|\delta C_{mn}(t)|$ is the main topic of the current paper, which can be obtained by the SPA,  where the key point is that the integral in $\delta C_{mn}^{(1)}(t)$ or $\delta C_{mn}^{(2)}(t)$ is dominated by the contributions near the extrema of $\varepsilon_k$, and the factor $e^{2i\varepsilon_kt}$ is replaced by a Gaussian by the Taylor expansion of $\varepsilon_k$ at the extrema $k_0$,  where $k_0$ in general is a saddle point. Then the integrals are calculable, and the scaling behaviors can be obtained.  
In summary, when $t$ is large enough, the integral (\ref{int1}) or (\ref{int2}) is approximately proportional to 
\begin{eqnarray}
	\small
	{\rm Re}\Bigg\{e^{i(2\varepsilon_{k_0}t+\varphi)}\int_{-\infty}^\infty dk (k-k_0)^q \exp\big[ib (k-k_0)^pt\big]\Bigg\}, \label{int3}
\end{eqnarray}
where  $\varphi$ and $b$ are trivial constants; $p$ is determined by the asymptotic behaviors  of the spectrum of the postquenched Hamiltonian in the vicinity of the saddle point, i.e., 
$\varepsilon_k\sim \varepsilon_{k_0}+b(k-k_0)^p$; $q$ is determined by the asymptotic behaviors of the term $\cos\zeta_k\sin^2\theta_k$ in Eq. (\ref{int1}) or the term  $\sin\zeta_k\sin(2\theta_k)$ in Eq. (\ref{int2}) as $k$ approaches $k_0$. 
 For example, for a  noncritical quench to the commensurate phase, in the vicinity of the saddle point $k=0$, 
$p=q=2$,  which eventually leads to a power law of $t^{-3/2}$.
For more details,  see the Appendix A of Ref. \onlinecite{quench1}, where more examples are given.  
 However, in Ref. \onlinecite{quench1} the analysis is restricted to $\delta C_{mn}^{(1)}(t)$, because the initial state is chosen as $(v_k(0),u_k(0))$=$(0,1)$.  
Generally, $|\delta C_{mn}^{(2)}(t)|$ follows the same scaling law of $|\delta C_{mn}^{(1)}(t)|$, however, for the critical quench studied in the current paper, 
we show that, in certain case, the scaling laws can be different. 
More importantly, the gap-closing property of $\varepsilon_k$ may substantially change the scaling behaviors,  
and new crossover phenomena can also be induced. 

 In the current paper,  in the calculations of  integrals (\ref{int1}) and (\ref{int2}), $m$ is set to be equal to $n$  if not explicitly stated.  The results do not have qualitative difference for $m\ne n$ if the distance between the sites $m$ and $n$ is short.

\section{Results}
\label{sec_short}
\subsection{Quech to a critical point}
When the prequeched Hamiltonian is noncritical and the postquenched Hamiltonian is critical,  the universal property of the  relaxation behavior of $|\delta C_{mn}(t)|$ is  the same as the corresponding noncritical quench (where  both the prequenched and postquenched Hamiltonians are  noncritical). 

\begin{figure}[htpb]
	\centering
	\includegraphics[width=0.9\columnwidth]{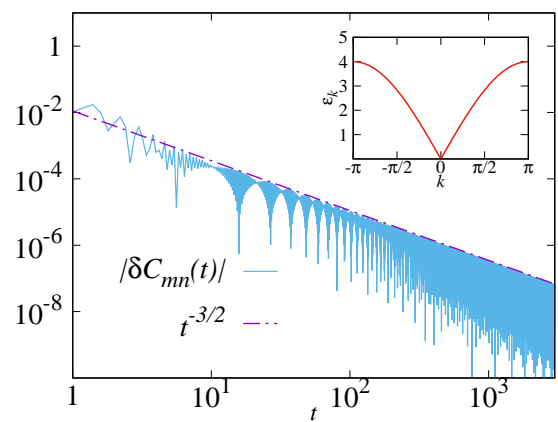}
	\caption{ Scaling behavior of a qench from ($h,\chi$)=(1.5, 2) to (1, 1); the subset is the energy spectrum of the postquenched Hamiltonian.}
	\label{n-c}
\end{figure}

Firstly, we consider the quench to a {\it critical commensurate phase}, i.e., the postquenched Hamiltonian has the parameters $(h=\pm1,  |\chi|<\sqrt{2})$.
A typical example is shown in Fig. \ref{n-c}, which is a quench from ($h,\chi$)=(1.5, 2) to (1, 1).
Here the scaling relation $ |\delta C_{mn}(t)| \sim t^{-3/2}$ is kept,  which is the contribution of the saddle point $k=\pi$.
Nearing this point, $\sin\zeta_k\sim k-\pi$ and $\sin\theta_k\sim k-\pi$ too, hence $q=2$ in Eq. (\ref{int3}); 
furthermore, $\varepsilon_k=\varepsilon_\pi+b(k-\pi)^2$, where $b$ is the second derivative of $\varepsilon_k$ at $\pi$, this gives $p=2$ in Eq. (\ref{int3}).
The integral (\ref{int3}) with $q=2$ and $p=2$  gives a $t^{-3/2}$ scaling.
It should be noted that $k=0$ is not a saddle point, it does not contribute a power-law scaling in the dynamics  but an exponential decaying, as we will show in the following example.

\begin{figure}[htpb]
	\centering
	\includegraphics[width=0.9\columnwidth]{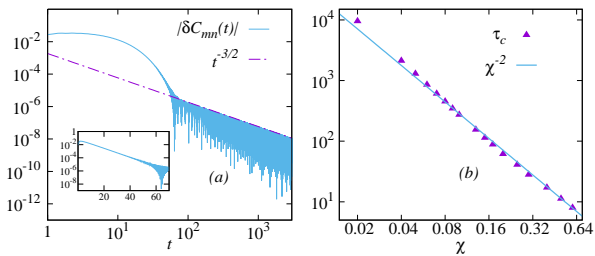}
	\caption{Crossover of a qench from ($h,\chi$)=(1.5, 2) to (1, 0.2); the subset is a quasi-logarithmic plot of the early time region.}
	\label{n-c2}
\end{figure}

When  $|\chi|<1$, especially when $\chi$ is small, it gives an interesting relaxation behavior; a typical example is shown in Fig. \ref{n-c2}(a). When $t$ is large enough, $|\delta C_{mn}(t)|$ is  dominated by the $t^{-3/2}$ scaling; however, in the early time region $(0,\tau_c)$,  $|\delta C_{mn}(t)|$ is dominated by an exponential decaying behavior. 
Such type of crossover  is owing to the competition between the saddle point $k=\pi$ and the gap-closing point $k=0$.
The saddle point $k=\pi$ leads to the scaling of $t^{-3/2}$, the reason is the same as the aforementioned case shown in Fig. \ref{n-c}. 
The exponential decaying in the early time region is owing to the contribution of the gap-closing point $k=0$.
It should be noted that although the  point $k=0$ is  not a saddle point, it is still an extreme point. 
The Taylor expansion of  $\varepsilon_k$ at this point should take the form
\begin{eqnarray}
	\varepsilon_k\sim \varepsilon_0+bk+\cdots,\\
	{\rm with} ~ b=\frac{d\varepsilon_k}{dk}|_{k\rightarrow 0^+},
\end{eqnarray}
because the first order derivative $b$ is not zero, the term $bk$ must have certain contribution to the integrals in Eqs. (\ref{int1}) and (\ref{int2}),
which is the origination of the exponential decaying behavior.  
Furthermore, because $\varepsilon_0=0$, in Eq. (\ref{int3}) the prefactor  $e^{i(2\varepsilon_0t+\varphi)}$ does not depend on time, therefore in such an  exponential decaying region,  there is no oscillation.
The crossover phenomenon is obvious when the value of $\chi$ is small, because in this case  the value of the $t^{-3/2}$ term is small, it can dominate the relaxation behavior only when the exponential term decays to much smaller value.  
The  crossover time $\tau_c$ is numerically determined, we find that it diverges with a power law 
\begin{eqnarray}
\tau_c\sim\chi^{-\kappa} \label{cross}
\end{eqnarray}
as $\chi\rightarrow 0$, where $\kappa\approx2.0$. This is demonstrated in Fig. \ref{n-c2}(b)

However, for the postquenched Hamiltonian with $\chi$ exactly at 0, i.e.,  a quench to the Luttinger liquid (which is also a critical phase), we find that $C_{mn}(t)$ does not evolve with time, this is easy to understand, because in this case, the Hamiltonian (\ref{shortfermion}) only has the hopping term,  the number of particles is conserved. 

\begin{figure}[htpb]
	\centering
	\includegraphics[width=1\columnwidth]{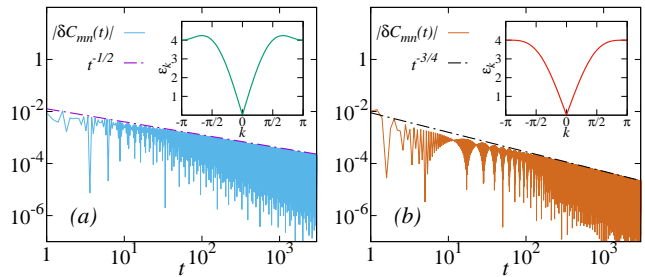}
	\caption{(a) scaling behavior of a quench from ($h,\chi$)=(1.5, 2) to a critical incommensurate phase with $(h,\chi)=(1,\sqrt{3})$; (b) scaling behavior of a qench from ($h,\chi$)=(1.5, 2) to $(1,\sqrt{2})$.  The subsets are the energy spectrum of the postquenched Hamiltonians.}
	\label{n-c3}
\end{figure}

We then consider the  quench to a {\it critical incommensurate phase}, i.e., the postquenched Hamiltonian has the parameters $(h=\pm1, |\chi|>\sqrt{2})$. 
A typical example is shown in Fig. \ref{n-c3}(a), which is a quench from $(h,\chi)$=(1.5,2) to $(1,\sqrt{3})$; the postquenched Hamiltonian has an additional saddle point $k_0=2\pi/3$ in the energy spectrum.  It is shown that in this case the relaxation scaling law $t^{-1/2}$ is kept, falling in the prediction of SPA. 

At last, we consider a quench from $(h,\chi)$=(1.5, 2) to $(1, \sqrt{2})$, where the postquenched Hamiltonian is at  the boundary between the critical commensurate phase and the critical incommensurate phase,  as shown in Fig. \ref{n-c3}(b);  it is shown that in this case the relaxation scaling law $t^{-3/4}$ is kept, consistent with  the theory of SPA.

\begin{figure}[htpb]
	\centering
	\includegraphics[width=1\columnwidth]{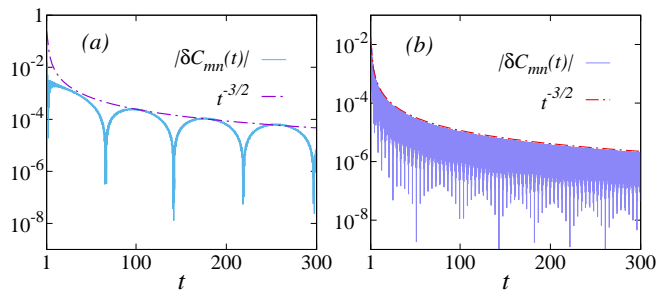}
	\caption{(a) dynamical relaxation of a qench from ($h,\chi$)=(1.5, 2) to (0.99,1); (b) dynamical relaxation of a qench from ($h,\chi$)=(1.5, 2) to $(1,1)$.}
	\label{n-cx}
\end{figure}

If the postquenched Hamiltonian is close to but not exactly at the critical point, then the period  becomes very large, as shown in Fig. \ref{n-cx}(a);  this is in sharp contrast to the  case that is exactly at the critical point, as shown in Fig. \ref{n-cx}(b). The reason is that the circular frequency  of the oscillation of $\delta C_{mn}(t)$ is  $\omega=2\varepsilon_{k_0}$, as shown in the prefactor of the integral of Eq. (\ref{int3});  therefore the circular frequency of $|\delta C_{mn}(t)|$ should be $\omega=4\varepsilon_{k_0}$.  For the quench with the postquenched Hamiltonian that is exactly at the critical point, the $k=0$ is not a saddle point, and the circular frequency $\omega=\omega_\pi=4\varepsilon_{\pi}=16$, therefore the period is $T=2\pi/\omega=\pi/8$;
 however, for the quench with the postquenched Hamiltonian that is close to but not exactly at the critical point,  there are two circular frequencies, in which $\omega_\pi=15.92$ and $\omega_0=0.08$; correspondingly, there are two periods $T_\pi=\pi/7.96$ and $T_0=25\pi$;  We can see that $T_0$ is much larger than $T_\pi$,  and the oscillation behavior is very similar to a beat. 

\subsection{Quench from a critical point}
\label{fromc}
If the prequenched Hamiltonian is at a critical point, the scaling behavior in the dynamics may change substantially, depending on the choosing of  both the prequenched Hamiltonian and the postquenched Hamiltonian. 

\begin{figure}[htpb]
	\centering
	\includegraphics[width=0.9\columnwidth]{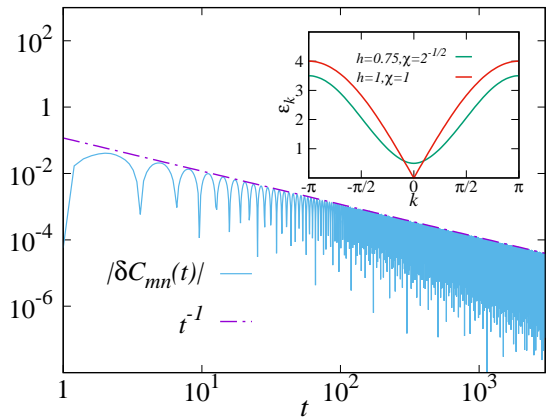}
	\caption{Scaling behavior of a quench from $(h,\chi)=(1,1)$ to $(0.75,1/\sqrt{2})$; the subset shows the energy spectrum.}
	\label{shortc-n1}
\end{figure}  
Firstly, we study the quench from a critical point to a commensurate phase. A typical example is shown in Fig. \ref{shortc-n1}, which is a quench from $(h,\chi)$=(1,1) to ($0.75,1/\sqrt{2}$). Here the scaling law 
of the relaxation behavior is $t^{-1}$, which is different from the $t^{-3/2}$ scaling of the corresponding noncritical quench. 
For an understanding of such result, we pay attention to Eq. (\ref{int2}),
in the vicinity of the saddle point $k=0$, which is also the gap-closing point of the 
prequenched Hamiltonian,  $\sin\zeta_k\sim 1$ and $\sin(2\theta_k)\sim k$,  this means $q=1$. 
Furthermore, according to SPA, the spectrum of the postquened Hamiltonian can be replaced by the Taylor expansion at the saddle point, thus $\varepsilon_k =\varepsilon_{0}+bk^2$, 
where $b=\varepsilon_{0}^{\prime\prime}$ is the second derivative of $\varepsilon_k$ at $k=0$, therefore $q=2$.
Substitute these values of $p$ and $q$  to the integral  (\ref{int3}), we get a power law of $t^{-1}$. 

In fact,  in the current case, the integral of  Eq. (\ref{int1}) should contribute  a $t^{-2}$ scaling, because in the vicinity of the saddle point $k=0$, 
 the term  $\cos\zeta_k\sin^2\theta_k \sim k^3$, i.e., $q=3$ for Eq. (\ref{int3}).
 However, similar analysis show that in the vicinity of the saddle point $k=\pi$, the integral (\ref{int1}) should contribute  a $t^{-3/2}$ scaling.
Therefore,   the decaying behavior of 	$|\delta C_{mn}(t)|$ satisfies a formula of the mixture of $t^{-1}$, $t^{-3/2}$, and $t^{-2}$; 
the $t^{-3/2}$ and  $t^{-2}$ terms decay faster than $t^{-1}$ term,  they are  overwhelmed by the $t^{-1}$ term, hence eventually the dynamical relaxation process  is dominated by the $t^{-1}$ scaling. 

\begin{figure}[htpb]
	\centering
	\includegraphics[width=0.9\columnwidth]{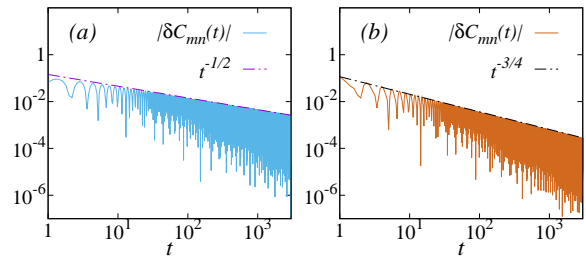}
	\caption{(a) scaling behavior  of a quench from $(h,\chi)=(1,1)$ to $(0.5,1/\sqrt{2})$; (b) scaling behavior of a quench from $(h,\chi)=(-1,1)$ to $(0.5,1/\sqrt{2})$.  }
	\label{shortc-n2}
\end{figure}

Secondly, we study the  quench from a critical point  to the boundary between the commensurate  and incommensurate phases,  in this case, the scaling law of the relaxation behavior may be changed or not, depending on the choosing of the prequenched (critical) Hamiltonian. As shown in Fig. \ref{shortc-n2}(a), a quench  from $(h,\chi)$=(1,1) to (0.5, $1/\sqrt{2}$) satisfies the scaling law $t^{-1/2}$,  which is different from the $t^{-3/4}$ scaling of the corresponding noncritical quench.  As to the reason of such change, it should be noted that  in the current case the integral (\ref{int1}) still gives a scaling of $t^{-3/4}$,  the $t^{-1/2}$ scaling  is given by (\ref{int2}).  
The reason is also related to the gap-closing of the energy spectrum of the prequenched Hamiltonian at $k=0$, which leads to $\sin\zeta_k\sim 1$, subsequently $\sin\zeta_k\sin(2\theta_k)\sim k$ in Eq. (\ref{int2}). Therefore, in Eq. (\ref{int3}) $q=1$; furthermore, because the postquenched Hamiltonian is at the boundary of the commensurate phase and incommensurate phase, 
the spectrum can be expanded as $\varepsilon_k=\varepsilon_0+\varepsilon_0^{\prime\prime\prime\prime}k^4$, because the second and third derivatives are zero, therefore $p=4$.
Substitute these values of $p$ and $q$ to Eq. (\ref{int3}), we get a power law of  $t^{-1/2}$.  
However, a quench  from $(h,\chi)$=(-1,1) to (0.5,$1/\sqrt{2}$) still satisfies  the power law of $t^{-3/4}$, this is shown in Fig. \ref{shortc-n2}(b). 
In this case, the gap-closing point of the spectrum of the  prequenched Hamiltonian is $k=\pi$, it does not interfere with the saddle point $k=0$ of the spectrum of the postquenched Hamiltonian, hence the power law of $t^{-3/4}$ is kept.

At last, we study the quench from a critical point to the incommensurate phase, in this case, the scaling law is not changed, for example, in a quench from $(h,\chi)$=(1,1) to $(0.25,1/\sqrt{2})$,
the scaling law $t^{-1/2}$  is still kept. The gap-closing point  $k=0$  of the prequenched Hamiltonian does not change the universal relaxation behavior of such a quench. The reason can be analyzed by a similar way. 

\begin{figure}[htpb]
	\centering
	\includegraphics[width=1\columnwidth]{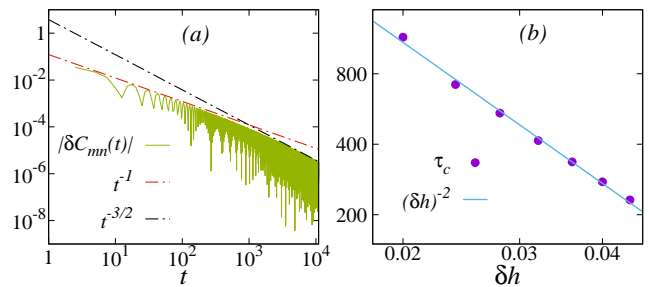}
	\caption{(a) crossover behavior of a quench from $(1+\delta h,\chi)=(1.02,1)$ to $(0.75,1/\sqrt{2})$; (b) crossover time $\tau_c$ versus $\delta h$.  }
	\label{c-n-cross}
	\end{figure}
If the prequenched Hamiltonian is very close to but not exactly at the critical point, crossover behavior between different scaling laws appears. 
Fig. \ref{c-n-cross} shows a typical example,   where the prequenched Hamiltonian is at ($1+\delta h,\chi$) = ($1.02, 1$)).  We can see that the $t^{-1}$ scaling behavior dominates when the time $t<\tau_c$, 
then it changes  to the scaling $t^{-3/2}$ for $t>\tau_c$.  
The reason for such crossover behavior is also  related to the gap closing of the  prequenched Hamiltonian: 
The width of the exponential term $\exp\big[ib (k-k_0)^pt\big]$ in Eq. (\ref{int3}), i.e., the effective range of $k$,  is $k_w\sim t^{-1/p}$; therefore, when $t$ is not large enough,   $k_w$  is obviously larger than $\delta h$, hence $\delta h $ is negligible, in this case, the spectrum of the prequenched Hamiltonian $\varepsilon_k \sim k$,
this leads to  $q=1$ in Eq. (\ref{int3}), therefore we get a $t^{-1}$ scaling. When $t$ is large enough, $k_w$ is very small,  subsequently $\delta h$ is comparable to $k$;  in this case, $q=2$, therefore we get a $t^{-3/2}$ scaling. We numerically determine the crossover time $\tau_c$,  
which also satisfies a power-law scaling $\tau_c\sim (\delta h)^{-\kappa}$. By the data fitting, we find that $\kappa\approx 2.0$, this is demonstrated in Fig. \ref{c-n-cross}(b).
It is a pity that currently we can not theoretically prove why the crossover exponent takes such value. 

\begin{figure}[htpb]
	\centering
	\includegraphics[width=1\columnwidth]{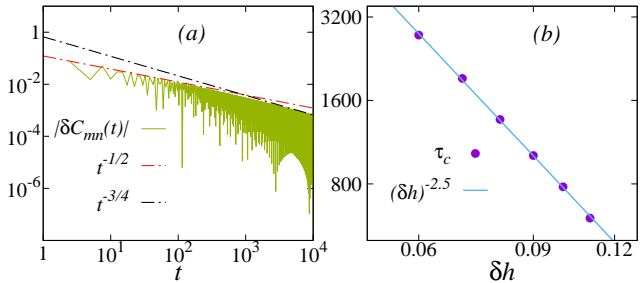}
	\caption{(a) crossover behavior of a quench from $(1+\delta h, \chi)=(1.1, 0.5)$ to $(0.5,1/\sqrt{2})$; (b) crossover time $\tau_c$ versus $\delta h$.  }
	\label{c-n-cross2}
\end{figure}
By the similar way, we investigate the quench from the vicinity of a critical point to the boundary between the commensurate  and incommensurate phases, a typical example is shown in Fig. \ref{c-n-cross2}(a) and the crossover time $\tau_c$ is shown in Fig. \ref{c-n-cross2}(b), where the crossover exponent is $\kappa\approx 2.5$.

\subsection{Quench from a critical point to another critical point}
\label{fromcc}
\begin{figure}[htpb]
	\centering
	\includegraphics[width=0.9\columnwidth]{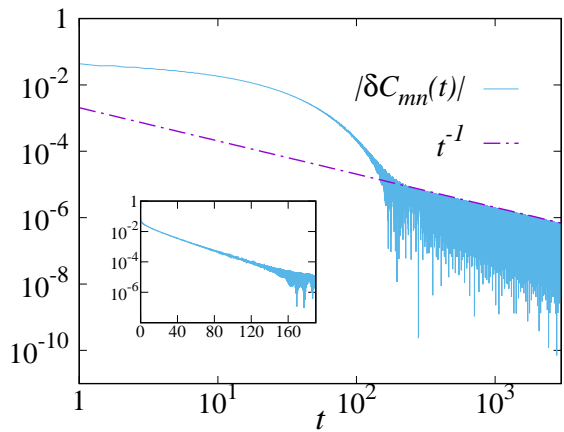}
	\caption{Crossover behavior of a quench from $(h,\chi)=(-1,1)$ to $(1,0.1)$; the subset is a quasi-log plot for the early time.}
	\label{shortc-c}
\end{figure}

An interesting question is, if both the prequenched and postquenched Hamiltonians are all at the critical points, then what happens?  In this case, we can find a mixture of the aforementioned results, for example, Fig. \ref{shortc-c} shows a quench from $(h,\chi)$=$(-1,1)$ to ($1,0.1$), we can see that the early-time dynamics in the region $(0,\tau_c)$ is dominated by an exponential decaying, and the long-time dynamics is dominated by a power-law of $t^{-1}$.  The crossover time $\tau_c$ also satisfies a scaling law $\tau_c\sim \chi^{-\kappa}$, with $\kappa\approx2.5$.

\subsection{Quench from a multicritical point}
\begin{figure}[htpb]
	\centering
	\includegraphics[width=1\columnwidth]{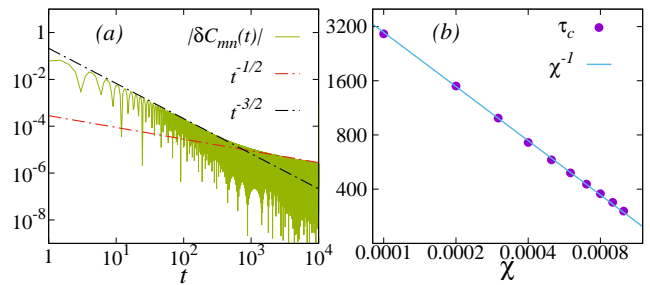}
	\caption{(a) crossover behavior of a quench from ($h,\chi$)=(1,0.0005) to ($0.75,1/\sqrt{2}$); (b) crossover time $\tau_c$ versus $\chi$.}
	\label{mc1}
\end{figure}  
In the phase diagram of Fig. \ref{pd}, there are two multicritical points, with $(h,\chi)=(\pm 1,0)$. In this section, we study the quench from such muliticritical point or the vicinity of such multicritical point.

If the prequenched Hamiltonian is exactly at a multicritical point, 
we find that  the scaling of the relaxation behavior does not change. 
Firstly, we investigate the quench from such multicriticl point to a commensurate phase, for example,  a quench from $(h, \chi)$=(1, 0) to the point $(0.75, 1/\sqrt{2})$ gives a scaling behavior of $^{-3/2}$. 
In this case, because $\chi$ is zero for the prequenched Hamiltonian,  then the integral in Eq. (\ref{int2}) is zero, $\delta C_{mn}(t)$ is completely determined by Eq. (\ref{int1}), 
where $\cos\zeta_k=1$ and $\sin^2\theta_k\sim k^2$, i.e., $q=2$ in Eq. (\ref{int3}), this leads to a $t^{-3/2}$ scaling behavior.  

Secondly, we investigate the quench from the multicritical point to the incommensurate phase and the boundary between the commensurate and incommensurate phases, we find the $^{-1/2}$ and $t^{-3/4}$ scaling behaviors, respectively. The theoretical analysis for these results can be performed similarly. 

However, if we set the prequenched Hamiltonian to be a point that is very close but not exactly at the multicritical point, we can find very interesting crossover phenomena.
For example, a quench from $(h, \chi)$=(1,0.0005) to the point $(0.75, 1/\sqrt{2})$ exhibits a crossover  between the $t^{-3/2}$ and $t^{-1/2}$ scaling behaviors, as shown in Fig. \ref{mc1}(a).
The crossover time $\tau_c$ versus different values of $\chi$ is shown in Fig. \ref{mc1}(b). $\tau_c$ satisfies a scaling law $\tau_c\sim \chi^{-\kappa}$, where $\kappa=-1$.
The condition for such type of crossover is that the value of $\chi$ should be very small (but not equal to zero), then the energy spectrum of the prequenched Hamiltonian $\varepsilon_k \sim k^2$ 
and subsequently in Eq. (\ref{int1}) the term $\cos\zeta_k\sim 1$, this leads to the $t^{-3/2}$ scaling behavior. Meanwhile, also because the energy spectrum of the prequenched Hamiltonian $\varepsilon_k \sim k^2$,
in Eq. (\ref{int2}) the term $\sin\zeta_k\sin\theta_k\sim \chi$, this leads to a $t^{-1/2}$ scaling. In summary, the scaling behavior of $|\delta C_{mn}(t)|$ in the current case takes the form 
\begin{eqnarray}
	|\delta C_{mn}(t)| = at^{-3/2}+bt^{-1/2},
\end{eqnarray}
where $a$ and $b$ are nonuniversal parameters.  Note that $b\sim \chi$ is very small;    
therefore, when the time is not long enough, the first term dominates; however, the first term decays faster than the second term, when the value of the first term is smaller than  the  second term, the second term begins to show the importance and eventually dominates scaling behavior.  It is obvious that the scaling for the first term to reach the second term is proportional to $t^{-3/2}/t^{-1/2}=t^{-1}$, therefore the crossover time $\tau_c$ satisfies $\tau_c^{-1}=\chi$, i.e., $\tau_c=\chi^{-1}$.  We can see that the mechanism of such type of crossover behavior is very different from the aforementioned crossover behaviors in Sections \ref{fromc} and \ref{fromcc}.

\begin{figure}[htpb]
	\centering
	\includegraphics[width=1\columnwidth]{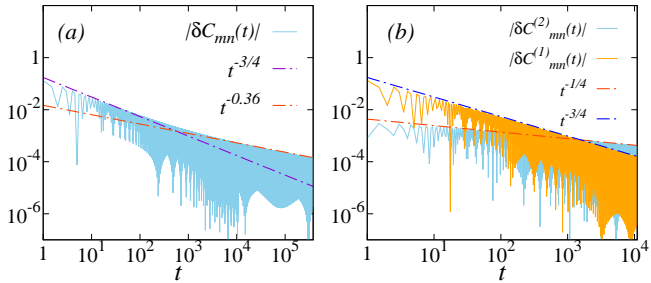}
	\caption{(a) Scaling behavior of $|\delta C_{mn}(t)|$ of a qench from ($h,\chi$)=(1,0.01) to ($0.5,1/\sqrt{2}$);  (b) scaling behaviors of $|\delta C^{(1)}_{mn}(t)|$ and $|\delta C^{(2)}_{mn}(t)|$ of the quench from ($h,\chi$)=(1,0.01) to ($0.5,1/\sqrt{2}$).}
	\label{mc2}
\end{figure}  
By the similar way, we study the quench from the vicinity of the multicritical point to the boundary between the commensurate and incommensurate phases, 
such as the quench from $(1,\chi)$ to $(0.5,1/\sqrt{2})$, with $\chi$ very small, we find a crossover from the  $t^{-3/4}$ scaling to the $t^{-1/4}$ scaling; here the $t^{-1/4}$ scaling comes from the integral (\ref{int2}),  which gives $q=0$ and $p=4$ for Eq. (\ref{int3}).  
Figure \ref{mc2}(a) shows a typical example of such type of crossover, we can see that in the early time, the relaxation is dominated by the $t^{-3/4}$ scaling, as we expected; however, in the later time that we have reached, it is dominated by a scaling of $t^{-0.36}$ but not $t^{-1/4}$, the reason is that in this region the term of $t^{-3/4}$ is not small enough, hence the relaxation is a mixture of $t^{-3/4}$ and $t^{-1/4}$, i.e., 
\begin{eqnarray}
	|\delta C_{mn}(t)|=at^{-3/4}+bt^{-1/4};
\end{eqnarray} 
in numerical sense, if the difference between $a$ and $b$ is not significant, the data may be fitted by a uniform scaling $t^{-\mu}$, with $-3/4<\mu<-1/4$. 
The $t^{-1/4}$ scaling can only dominate in very large time, this makes it very difficult to numerically determine the crossover time,  because when the time is too large, the revival phenomenon\cite{revival} may appear for a finite-size system.  In Fig. \ref{mc2}(a),  we have used a very large system size $L=10^{6}$, and the largest time shown  is $t=3.8\times 10^5$, above which the revival phenomenon appears.
For a further understanding of the question, we calculate the integrals of Eq. (\ref{int1}) and Eq. (\ref{int2}) respectively, the results are presented in Fig. \ref{mc2}(b), 
where the $t^{-3/4}$ scaling of $	|\delta C^{(1)}_{mn}(t)|$ and the $t^{-1/4}$ scaling of $	|\delta C^{(2)}_{mn}(t)|$ are shown very clearly.   
The scaling for the $t^{-3/4}$ term to reach the $t^{-1/4}$ term is proportional to $t^{-3/4}/t^{-1/4}=t^{-1/2}$,  and $b$ is proportional to $\chi$, therefore the crossover time $\tau_c$ satisfies $\tau_c^{-1/2}=\chi$, i.e., $\tau_c=\chi^{-2}$.  

\section{Summary and Discussion}
\label{con}
In summary, we have studied the universal real-time dynamical relaxation behaviors of a quantum XY chain after a 	``critical quench", where the  prequenched Hamiltonian, or the postquenched Hamiltonian, or both of them are at critical points of  equilibrium quantum phase transitions. 
Generally, a quench to a critical point does not change the universal power-law scaling behavior but may lead to a crossover between the exponential decaying behavior and the power-law scaling behavior.  
A quench from a critical point may change the power-law scaling behaviors $t^{-3/2}$ and $t^{-3/4}$ in the noncritical quenches to $t^{-1}$ and $t^{-1/2}$, respectively.
Furthermore, if the prequenched Hamiltonian is set to be the vicinity of a critical point, we can find interesting crossover between different scaling behaviors.
Similar questions are also studied in the quench from a multicritical point, we find crossover behaviors originating from a new mechanism, where new crossover exponent is found.

All  the new results in the critical quench are related to the gap-closing properties of the critical points, more specifically, at the critical point, 
$\varepsilon_k\sim k$,  this shelters the asymptotic behavior of $\sin\zeta_k$ or $\sin\theta_k$ in the integrals of (\ref{int1}) and (\ref{int2}) as $k$ approaches zero, 
 and subsequently leads to different scaling behaviors. 
 It is obvious that if the dispersion is different, it will lead to different scaling behaviors in the critical quench; the quench from the vicinity of the multicritical point, which has a dispersion  $\varepsilon_k\sim k^2$,   is a typical example.  It is an interesting question to study the relaxation behaviors after  a  critical quench in other integrable  and nonintegrable systems, such as the periodically driven system\cite{pbc}, the aperiodically driven  system\cite{apbc},  the stochastic driven system\cite{stoch}, the noise driven systems\cite{noise}, and so forth.  In fact, nonequilibrium dynamics involving critical states also lead to other interesting physics\cite{ding2020,Jafari2021,Yin2016,Yin2016a}.

\section*{Acknowledgment}
We thanks Aamir Ahmad Makki for valuable discussion; this work is supported by the National Science Foundation of China (NSFC) under Grant Numbers 11975024 and the Anhui Provincial Supporting Program for Excellent Young Talents in Colleges and Universities under Grant No. gxyqZD2019023.

\end{document}